\documentclass[aps,prl,superscriptaddress,showpacs,twocolumn]{revtex4}
\usepackage{graphicx}
\usepackage{amsmath}
\usepackage[version=3]{mhchem}

\usepackage[usenames]{color}
\definecolor{BrickRed}{cmyk}{0,0.89,0.94,0.28}
\definecolor{MidnightBlue}{cmyk}{0.98,0.13,0,0.43}
\definecolor{DarkGreen}{rgb}{0,0.7,0.1}
\newcommand{\add}[1]{#1}

\usepackage{footmisc}
\DefineFNsymbols{mySymbols}{{\ensuremath \ast}{\ensuremath{\dagger}}\S{\ensuremath \&}\P
   *{**}{\ensuremath{\ddagger}}}
\setfnsymbol{mySymbols}

\begin{document}


\title{
Glucagon stop-go kinetics supports\\a monomer-trimer fibrillation model.}


\author{Andrej Ko{\v s}mrlj}
\email{andrej@physics.harvard.edu}
\affiliation{Harvard University, Department of Physics, 17 Oxford Street, Cambridge, MA 02138, USA}

\author{Pia Cordsen}
\affiliation{Copenhagen University, Niels Bohr Institute, CMOL, Blegdamsvej 17, DK-2100 Copenhagen, Denmark}

\author{Anders Kyrsting}
\thanks{Now at {\it University of Cambridge, Department of Chemical Engineering and Biotechnology, Cambridge CB2 3RA, UK}}
\affiliation{Copenhagen University, Niels Bohr Institute, CMOL, Blegdamsvej 17, DK-2100 Copenhagen, Denmark}

\author{Daniel E. Otzen}
\affiliation{Copenhagen University, Niels Bohr Institute, CMOL, Blegdamsvej 17, DK-2100 Copenhagen, Denmark}
\affiliation{Interdisciplinary Nanoscience Center (iNANO), Department of Molecular Biology and Genetics, Aarhus University, Gustav Wieds Vej 14, DK-8000 Aarhus C, Denmark}

\author{Lene B. Oddershede}
\email{oddershede@nbi.dk}
\affiliation{Copenhagen University, Niels Bohr Institute, CMOL, Blegdamsvej 17, DK-2100 Copenhagen, Denmark}

\author{Mogens H. Jensen}
\email{mhjensen@nbi.dk}
\affiliation{Copenhagen University, Niels Bohr Institute, CMOL, Blegdamsvej 17, DK-2100 Copenhagen, Denmark}


\date{\today}

\begin{abstract}
We investigate {\it in vitro} fibrillation kinetics of the hormone peptide glucagon at various concentrations using confocal microscopy \add{ and determine the glucagon fibril persistence length $60 \mu\textrm{m}$}. At all concentrations we observe that periods of \add{individual} fibril growth are interrupted by periods of stasis.  The growth probability is large at high and low concentrations and is reduced for intermediate glucagon concentrations. To explain this behavior we propose a simple model, where fibrils come in two forms, one built entirely from glucagon monomers and one entirely from glucagon trimers. The opposite building blocks act as fibril growth blockers, and this generic model reproduces \add{experimental behavior} well.
\end{abstract}


\maketitle

Misfolding and aggregation of peptides and proteins into fibrils are the hallmarks of around 40 human diseases \cite{selkoe03,chiti06}.
Understanding the fibrillation process of one protein may provide a generic mechanistic insight useful for understanding fibrillation of a class of proteins.
In this paper we focus on the protein glucagon, which is a 29 amino acid residue hormone peptide, that upregulates blood sugar levels.
It is an important pharmaceutical molecule, which is used to treat diabetic patients in situations of acute hypoglycemia \cite{drucker05,jiang03}.
As obesity and the number of diabetic patients is increasing, this drug becomes more and more relevant.
The active state of glucagon is the monomer, but during pharmaceutical production the peptide has a high tendency to misfold  and aggregate into fibrils devoid of biological function \cite{pedersen10}.
When glucagon is solubilized, it can be found in two states, which produce glucagon fibrils of different morphologies.
Below a concentration of 1 mg/mL, glucagon is predominantly found in an unstructured monomeric state, while above 1 mg/mL glucagon form associated states such as trimers and other oligomers \cite{gratzer69,formisano77,johnson78,wagman80,svane08}.
The monomer and oligomer precursor states lead to twisted and non-twisted fibrils, respectively \cite{pedersen06,andersen07,andersen10}. 
Experiments suggest that at high glucagon concentrations, the monomeric species are not incorporated into fibrils \cite{svane08} and the growth of twisted fibrils is inhibited \cite{andersen07}.

Fibrillation of proteins and peptides is typically followed in bulk using the fibril-binding fluorescent dye Thioflavin T (ThT).
While ThT-based fibrillation kinetics can provide highly valuable information on the mechanisms of fibrillation \cite{cohen13}, studies of the growth of individual fibrils can also yield important insights.
This information is provided by techniques such as Total Internal Reflection Fluorescence Microscopy (TIRFM) and Confocal Microscopy (CM).
In TIRFM the observation depth is $\sim$150 nm while with CM it is $\sim$500 nm.

Previously, \add{we have studied} growth of individual glucagon fibrils in real-time using TIRFM \cite{borg10} at \add{one fixed} glucagon concentration.
In that study, fibril growth was found to be interrupted by periods of stasis, and the statistics of growth and stasis durations were well described by a Poissonian process. This dynamic behaviour was denoted {\it stop-go} kinetics. Switching rates between the growing and arrested states suggested the probability of being in the growing state to be $\sim$1/4.
To explain this value, a Markovian four-state model of fibril growth was proposed.
The model predicted that the growth probability is \emph{independent} of the glucagon concentration.
This is in contrast to our findings since here we demonstrate that the fibril growth probability \emph{does} depend on the glucagon concentration.

\begin{figure}[t!]\centering
\includegraphics[width=.45\textwidth]{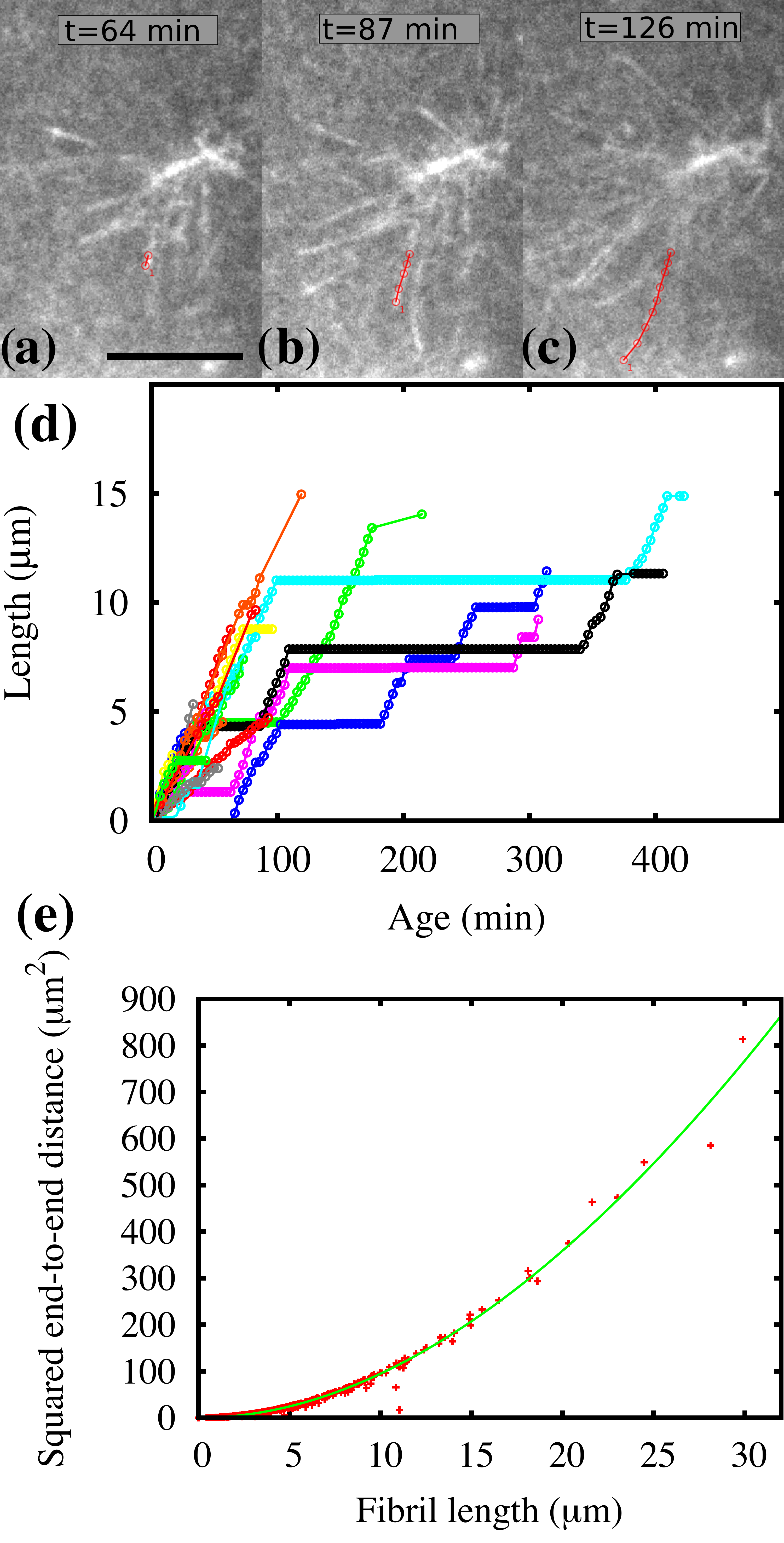}
\caption{
(Color)
(a-c): Confocal
microscopy images of glucagon fibrils with initial concentration 3 mg/mL in
aqueous buffer (50 mM glycine HCl, pH 2.5) at three consecutive times: 64, 87
and 126 mins after the onset of fibrillation.  Scale bar shows 5 $\mu$m.
Each circle represents a data point and the red line represents the cumulated
tracked positions of the growing fibril end.  (d) Growth of 20 fibrils at 
\add{the glucagon} concentration of 3 mg/mL.
Plateaus correspond to arrested states while fibrils elongate outside the
plateaus. 
\add{(e) The average end-to-end-distance squared ($\left< R_{ee}^2 \right>$) as a function of fibril length. The solid green line is obtained by fitting Eq.~(\ref{eq:persistence_length}) to combined experimental data (red points) from all glucagon concentrations. The persistence length of fibrils is returned by the fit as  $60 \pm 2 \mu\textrm{m}$.}
} \label{fig.tracks}
\end{figure}

\add{Here we significantly expand our previous work~\cite{borg10} by monitoring} fibril kinetics over a wide range of glucagon concentrations\add{. The advantage of this approach is that it allows us to sample conditions spanning} different precursor states of glucagon\add{, i.e. monomers or trimers, leading} to twisted or straight fibrils, respectively. Fibrils were labeled with the fluorescent dye ThT and monitored using a confocal microscope with an Argon laser.
On freshly plasmated glass plates we observed a volume of $\sim$40$\times$40$\times$0.5 $\mu$m$^3$.
For each of the five different initial glucagon concentrations (1.5, 3, 6, 10 and 15 mg/mL), a minimum of two experiments were conducted in aqueous buffer (50 mM glycine HCl, pH 2.5).
The time interval between captured frames was 3.3 mins and the total observation time of each experiment was about three days.
When fibrils grew along the surface we tracked their length as a function of time. Sample images of real time growth of an individual fibril are shown in Fig. \ref{fig.tracks}(a-c).
\add{The observed growing fibrils are relatively straight and their persistence length $\ell_p$ can be extracted by comparing the geometric distance between fibril ends $R_{ee}$ to the fibril length $L$. For semi-flexible fibrils the average end-to-end distance is expected to be~\cite{doiB}
\begin{equation}
\left< R_{ee}^2 \right> = 2 \ell_p L - 2 \ell_p^2 \left[ 1 - e^{-L/\ell_p}\right],
\label{eq:persistence_length}
\end{equation}
which agrees extremely well with experimental data (Fig.~\ref{fig.tracks}e). The fitting of equation above to experimental data provides a persistence length $\ell_p = 60 \pm 2 \mu\textrm{m}$. Note that this is of the same order as the persistence length of actin filaments \mbox{($\sim\!20 \mu\textrm{m}$)}~\cite{gittes93}, while much smaller than the persistence length of microtubules \mbox{($\sim\!5,000 \mu\textrm{m}$)}~\cite{gittes93}, and larger than the persistence lengths of DNA \mbox{($\sim\!50\textrm{nm}$)}~\cite{hagerman88} and amyloid fibrils \mbox{($0.1$--$4\mu{\textrm{m}}$)}~\cite{akker11}.
}
\begin{figure}[t]\centering
  \includegraphics[width=.45\textwidth]{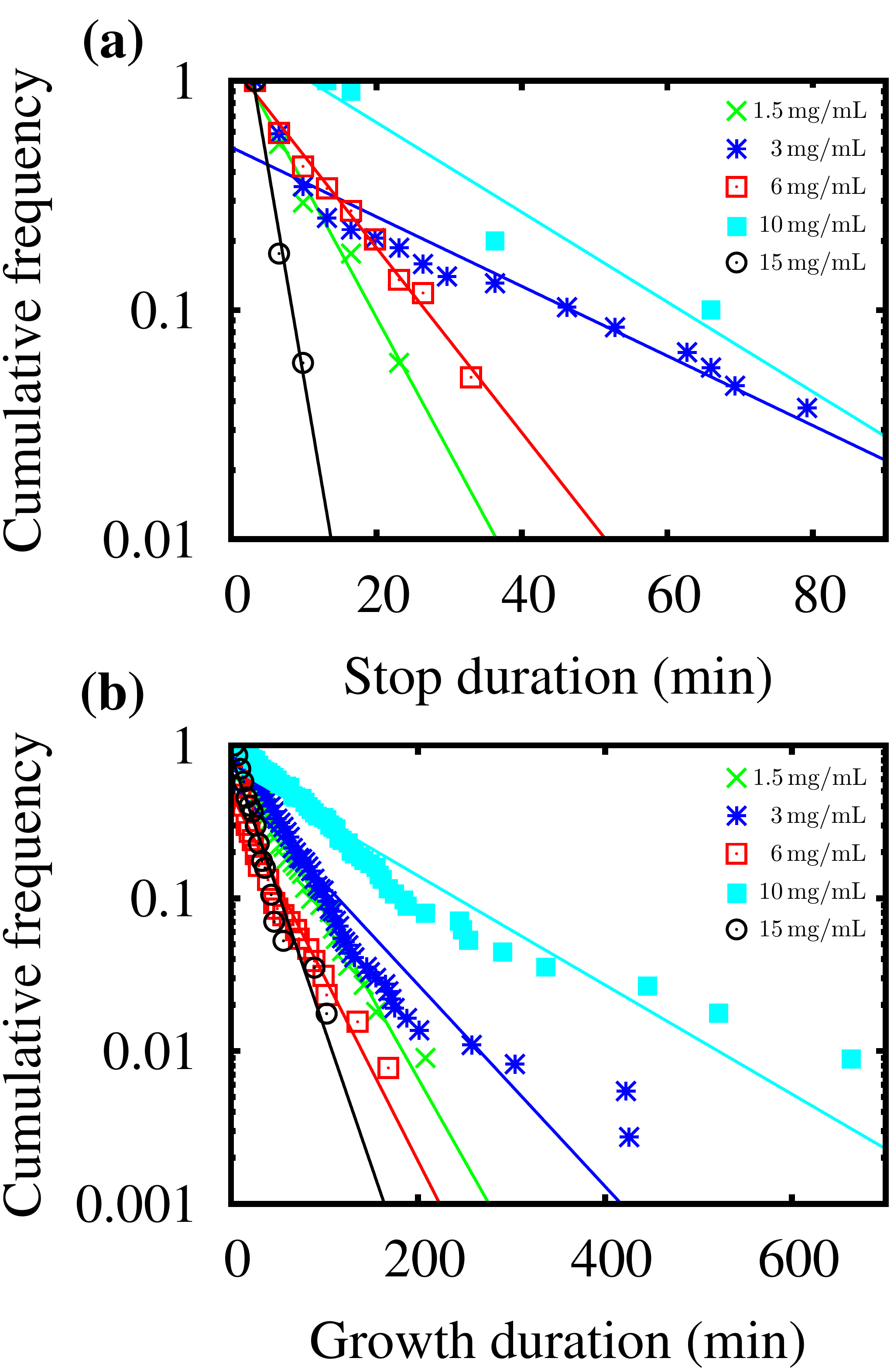}
\caption{
(Color)
\add{Distributions of stop (a) and growth (b) durations for fibrils grown at various glucagon concentrations.}
Straight lines indicate linear fits to the cumulative data.
Three extremely long pauses were removed from the 3 mg/mL sample.
}
\label{stopandgosammen1}
\end{figure}

\begin{figure*}[t]\centering
 \includegraphics[width=0.9\textwidth]{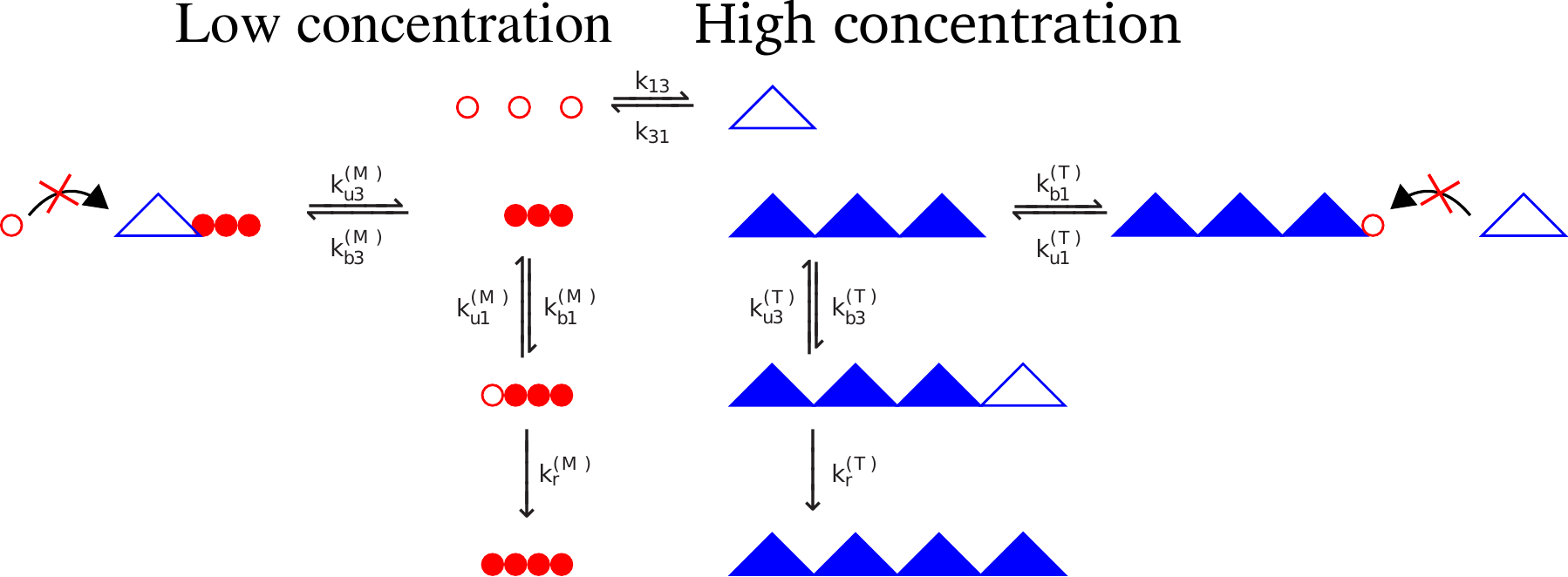}
\caption{
(Color online)
Schematic overview of the glucagon growth model showing (upper part) monomer-trimer equilibrium and (lower part) fibrillation process.
Glucagon monomers are in equilibrium with glucagon trimers.
Elongation of a fibril is a two-step process, which can be interrupted by binding of the other oligomer.
Fibrils consist of either monomers or trimers but never a combination of the two.
A glucagon trimer (monomer) can bind to a growing fibril end and then dissociate or elongate the fibril after conformational rearrangement.
Filled triangles (circles) symbolize trimers (monomers) bound irreversibly to a fibril, while hollow triangles (circles) mean unbound trimers (monomers).
A glucagon monomer (trimer) can also bind to a growing fibril end, but in this arrested state it prevents further attachment of glucagon trimers (monomers).
}
\label{fig.model}
\end{figure*}

\add{By inspecting the time courses of fibril lengths (Fig.~\ref{fig.tracks}d),  }
we find that at all glucagon concentrations the fibril growth is characterized by periods of growth (go
state) interrupted by periods of stasis (stop state). The stop states are seen as plateaus, where the fibril
does not elongate. \add{As seen in our previous work~\cite{borg10}, the} distributions of the stop and go event durations (displayed in Fig.~\ref{stopandgosammen1}) follow exponential distributions and are fitted to the form $f(x)=a\cdot\exp(-k\cdot t)$. A fibril leaves the go state at rate \add{$k_{\textrm{g}\rightarrow \textrm{s}}$} given by growth durations (Fig. \ref{stopandgosammen1}a) and leaves the stop state at rate \add{$k_{\textrm{s}\rightarrow \textrm{g}}$} given by stop durations (Fig. \ref{stopandgosammen1}b). \add{Both switching rates depend on the glucagon concentration. The access to kinetic data at different glucagon concentrations allows us to develop a model for glucagon's fibrillation.}

The analytical models for the kinetics of fibril growth \add{were} initiated with the Oosawa model \cite{oosawaB} and further elaborated to include hydrolysis and breakage of fibrils \cite{knowles09,arosio12}.
Our model is an extension of the Oosawa model, which includes both monomers and trimers as basic building blocks for fibrils.

To explain \add{the intermittent fibril growth} behavior we propose a model sketched in Fig.~\ref{fig.model}. In the bulk solution glucagon monomers are in equilibrium with glucagon trimers and these two components give rise to twisted and non-twisted fibrils, respectively. Successive binding of glucagon monomers to the twisted fibril end corresponds to the growing state, while binding of trimers to the twisted fibril end prevents further growth until the trimer is detached. During this time the twisted fibril appears to be in the arrested state. The opposite is true for non-twisted fibrils, which are formed from glucagon trimers, while glucagon monomers inhibit their growth.

In the mean field approximation, the fibril growth probability can be expressed in terms of the model rate constants and compared to the experimentally observed growth probabilities.
The growth probability predicted by the model is calculated by considering the average time spent in the growing or arrested state as outlined below.
In a bulk solution glucagon is in equilibrium between monomers (M) of concentration $[G]$ and trimers (T) of concentration $[G_3]$ with the equilibrium constant
\begin{align}
K_0^2 = \frac{[G]^3}{[G_3]} =\frac{k_{31}}{k_{13}}
\end{align}
and the total glucagon concentration $[G_{\textrm{tot}}]=[G]+3[G_3]$. As mentioned before at low (high) glucagon concentrations, i.e., $[G_{\textrm{tot}}] \ll K_0$  ($[G_{\textrm{tot}}] \gg K_0$), glucagon is predominantly in the monomer (trimer) state.

For the free growing twisted fibril end it takes on average the time $(k^{(M)}_{b1} [G] + k^{(M)}_{b3} [G_3])^{-1}$ before the glucagon monomer or trimer binds to the tip. This occurs with probabilities  $p^{(M)}_1$ or $p^{(M)}_3$ respectively, where
\begin{align}
p^{(M)}_1= \frac{k^{(M)}_{b1} [G]}{(k^{(M)}_{b1} [G] + k^{(M)}_{b3} [G_3])}=1-p^{(M)}_3.
\label{eq:monomer_binding_probability}
\end{align}
If a glucagon monomer is bound to the growing twisted fibril end, it takes on average the time $(k^{(M)}_{u1} + k^{(M)}_r)^{-1}$ for the glucagon monomer to unbind with probability $p^{(M)}_{1u}$ or to undergo conformational rearrangement and form a longer fibril with probability $p^{(M)}_{1g}$, where
\begin{align}
p^{(M)}_{1u} = \frac{k^{(M)}_{u1}}{k^{(M)}_r + k^{(M)}_{u1}}=1-p^{(M)}_{1g}.
\end{align}
The average time $\tau^{(M)}_1$ for a monomer to bind and subsequently either unbind or undergo conformational rearrangement to elongate the twisted fibril is
\begin{align}
\tau^{(M)}_{1} &= \frac{1}{(k^{(M)}_{b1} [G] + k^{(M)}_{b3} [G_3])} + \frac{1}{(k^{(M)}_{u1}+k^{(M)}_r)},
\end{align}
while the average time $\tau^{(M)}_3$  for the binding and unbinding of a glucagon trimer is
\begin{align}
\tau^{(M)}_3 &= \frac{1}{(k^{(M)}_{b1} [G] + k^{(M)}_{b3} [G_3])} + \frac{1}{k^{(M)}_{u3}}.
\label{eq:trimer_time}
\end{align}

We define the growth probability $p^{(M)}_G$ as the expected average fraction of time the twisted fibril spends in the growing state:
\begin{align}
p^{(M)}_G = \frac{p^{(M)}_1 p^{(M)}_{1g} \tau^{(M)}_{1}}{p^{(M)}_1 \tau^{(M)}_1 + p^{(M)}_3  \tau^{(M)}_{3}}.
\end{align}

Similarly, we can analyze the dynamics of the growing non-twisted fibrils, which are formed from glucagon trimers. The growth probability for non-twisted fibrils is then
\begin{align}
p^{(T)}_G = \frac{p^{(T)}_3 p^{(T)}_{3g} \tau^{(T)}_{3}}{p^{(T)}_1 \tau^{(T)}_1 + p^{(T)}_3  \tau^{(T)}_{3}},
\end{align}
where all quantities are defined in analogous way as above for the twisted fibrils. However, in this case the role of glucagon monomers and trimers is reversed, i.e., in Eqns.~(\ref{eq:monomer_binding_probability}-\ref{eq:trimer_time}) above one should replace $(M)$ with $(T)$ and make the $1 \leftrightarrow 3$ substitutions to obtain the relevant quantities. Since the number of twisted and non-twisted fibrils is proportional to the number of glucagon monomers and trimers, respectively, the probability $p_G$ that the randomly chosen fibril is found in the growing state is
\begin{align}
p_G = \frac{p^{(M)}_G [G] + p^{(T)}_G [G_3]}{[G] + [G_3]}.
\label{eq:model_growth_probability}
\end{align}

It is possible to derive the exact expression for the growth probability above in terms of the rate constants and the total glucagon concentration, but for simplicity we present only the asymptotic regimes at low and high glucagon concentration.
At low glucagon concentration, $[G_{tot}] \ll K_0$, the majority of glucagon is in the monomeric state. 
The slow time scales correspond to binding of glucagon monomers or trimers to the fibril ends and the growing probability for twisted fibrils is 
\begin{align}
p^{(M)}_{G} \approx \frac{k^{(M)}_r}{(k^{(M)}_{u1}+k^{(M)}_r)} \left[ 1 -\frac{k^{(M)}_{b3}}{k^{(M)}_{b1}}\left(\frac{[G_{tot}]}{K_0}\right)^{2} \right].
\end{align}
There are only a small number of non-twisted fibrils, whose growth is further suppressed by binding of glucagon monomers
\begin{align}
p^{(T)}_{G}  \approx \frac{k^{(T)}_{b3}k^{(T)}_r}{k^{(T)}_{b1}(k^{(T)}_{u3}+k^{(T)}_r)}\left(\frac{[G_{tot}]}{K_0}\right)^2.
\end{align}
The fibril  growth probability is thus approximately 
\begin{align}
p_{G}^{\textrm{low}} \approx \frac{p^{(M)}_{G} [G]}{([G]+[G_3])},
\label{low}
\end{align}
where $[G]/([G]+[G_3]) \approx 1- [G_\textrm{tot}]^2/K_0^2$.

At high glucagon concentrations, $[G_{tot}] \gg K_0$, most  of the glucagon is in the trimeric state.
The binding events are fast because of the large concentration of glucagon trimers and the slow time steps are the unbinding and conformational reconfiguration. The growth probability of non-twisted fibrils is approximately
\begin{align}
p_G^{(T)} \approx \frac{k^{(T)}_r}{(k^{(T)}_{u3}+k^{(T)}_r)}-\frac{k^{(T)}_{b1}k^{(T)}_r}{k^{(T)}_{b3}k^{(T)}_{u1}}\left(\frac{3K_0}{[G_{tot}]}\right)^{2/3}.
\end{align}
There are only a small number of twisted fibrils, whose growth is further suppressed by binding of glucagon trimers
\begin{align}
p_G^{(M)} \approx \frac{k^{(M)}_{b1}k^{(M)}_r k^{(M)}_{u3}}{k^{(M)}_{b3}(k^{(M)}_{u1} + k^{(M)}_r)^2}\left(\frac{3K_0}{[G_{tot}]}\right)^{2/3}.
\end{align}
The fibril  growth probability is thus approximately 
\begin{align}
p_{G}^{\textrm{high}} \approx \frac{p^{(T)}_{G} [G_3]}{([G]+[G_3])},
\label{high}
\end{align}
where $[G_3]/([G]+[G_3]) \approx 1- \left(3 K_0/[G_\textrm{tot}]\right)^{2/3}$.

At intermediate glucagon concentrations, $[G_{tot}] \sim K_0$, there is a mix of twisted and non-twisted fibrils whose growth is suppressed due to binding of the opposite glucagon components. 

\begin{figure}[t]
 \includegraphics[width=.45\textwidth]{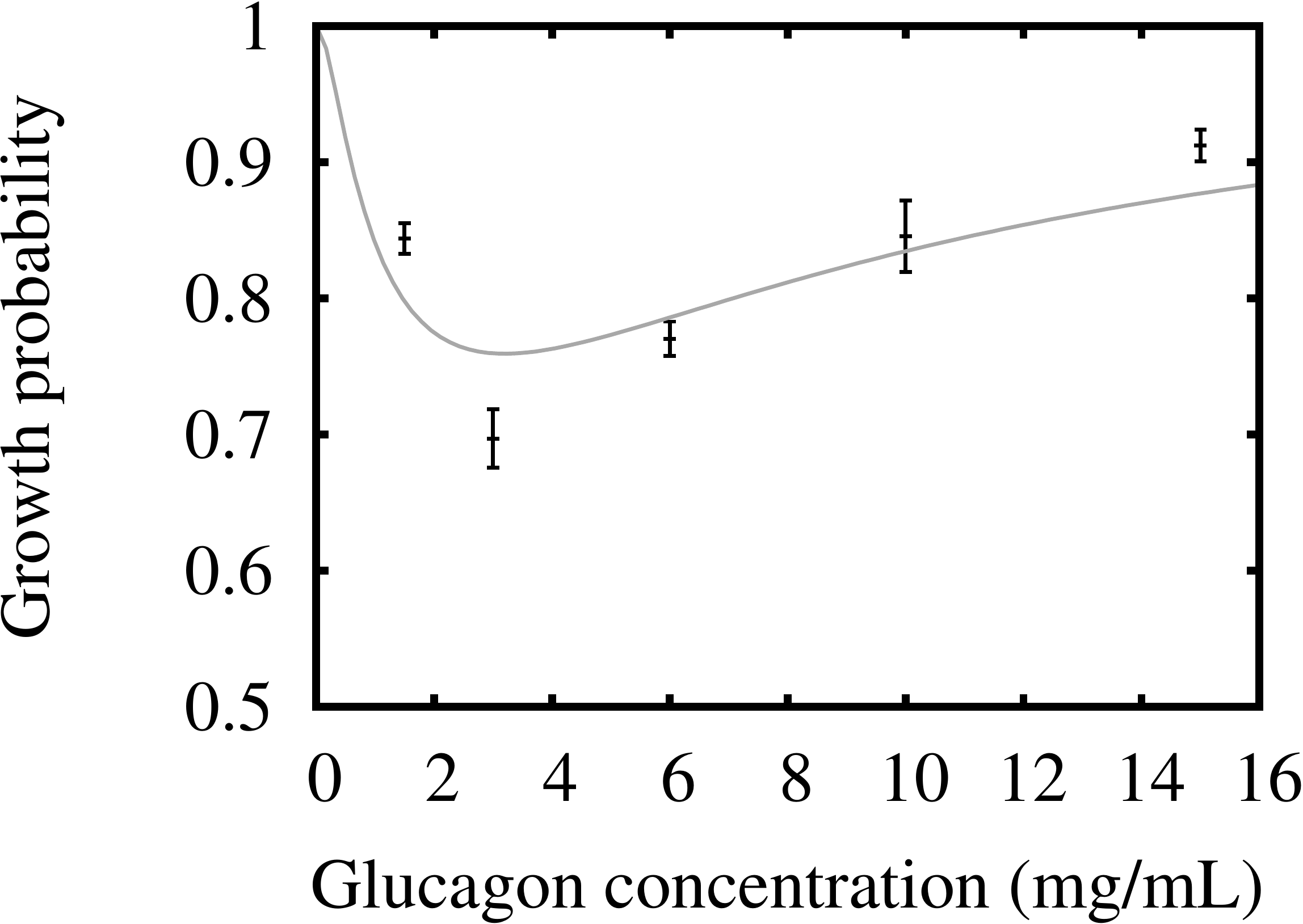}
\caption{
\add{Experimentally measured} fibril growth probabilities (black bars) calculated from Eq.~(\ref{eqn:pg}). 
The error bars in experimentally observed growth probabilities are calculated from uncertainties in \add{the switching rates between the stop and go states} from fitting in Figs. \ref{stopandgosammen1}(a-b).
The grey line shows a fit of the model in Eq.~(\ref{eq:model_growth_probability}) to the experimental data. 
The fit gives $K_0 \approx 1.3$ mg/mL.
}
\label{stopandgosammen2}
\end{figure}

The probability that at any moment a given fibril is in the growing state can be determined from the experimental \add{switching rates between the stop and go states} as
\add{
\begin{align}\label{eqn:pg}
p_G&=\frac{k_{\textrm{s}\rightarrow \textrm{g}}}{k_{\textrm{s}\rightarrow \textrm{g}} + k_{\textrm{g}\rightarrow \textrm{s}}}.
\end{align}
}
The measured fibril growth probabilities at different glucagon concentrations, given by Eq.~(\ref{eqn:pg}), are displayed in Fig.~\ref{stopandgosammen2} as black bars and are seen to qualitatively agree with the model behavior described above, which is plotted as a full line.
We notice that fibril growth probabilities are large at high and low glucagon concentrations, while they are smaller at intermediate glucagon concentrations ($\sim$3 mg/mL).
The value of $K_0 \approx 1.3$ mg/mL obtained from the fit is in accordance with previous studies of glucagon monomer-trimer equilibrium \cite{gratzer69,formisano77,johnson78,wagman80,svane08}.
A previous study of glucagon fibrillation at a very low concentration (0.25 mg/mL) found the growth probability to be $\sim$1/4 \cite{borg10}, which is smaller than the growth probabilities observed in our experiments (Fig.~\ref{stopandgosammen2}).
We speculate that in that study, fibril seeds grown at a higher glucagon concentration \add{could} bias the distribution of fibrils towards trimeric fibrils and hence result in a lower growth probability than predicted by our equilibrium model.

\begin{figure}[t]
 \includegraphics[width=.45\textwidth]{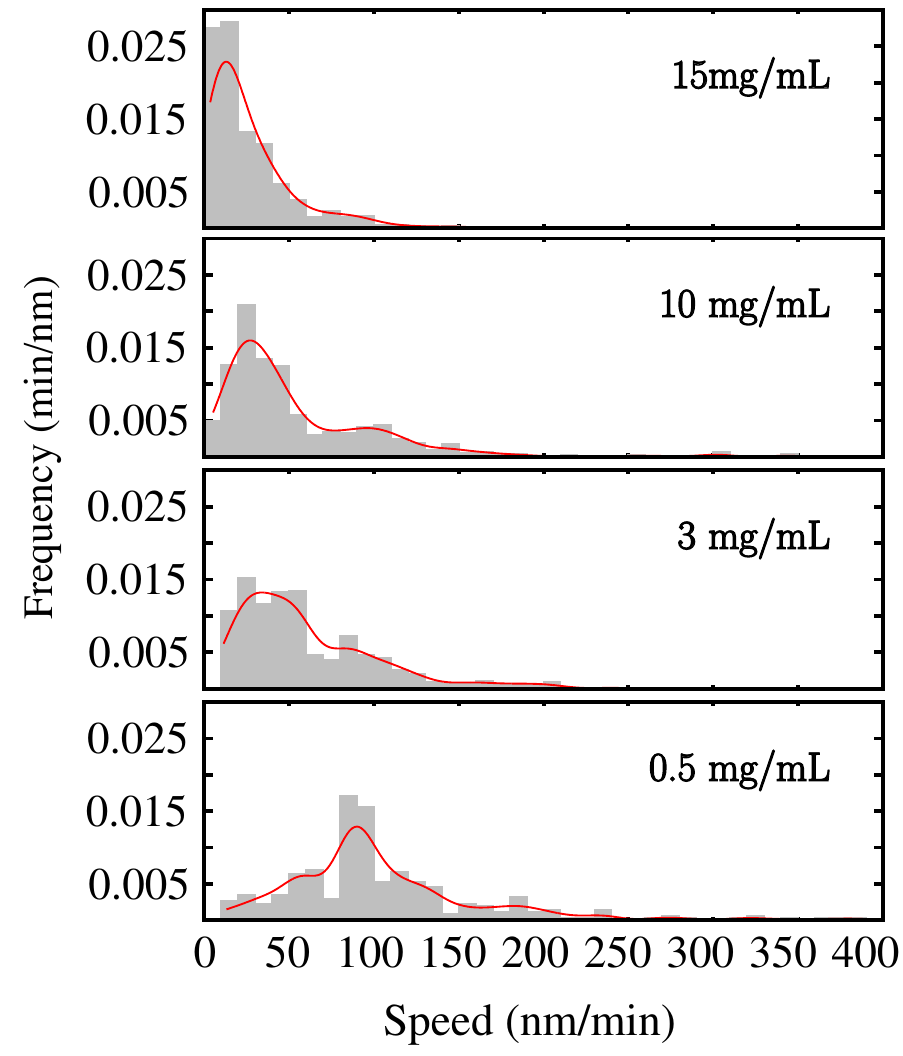}
\caption{
\add{(Color online) Growth speed distributions of glucagon fibrils at various glucagon concentrations (displayed in top-right corners)  are presented as histograms (grey boxes of width 10 nm/min). With red solid line we plotted an approximate distribution, where each experimental data point is represented as a Gaussian distribution with a fixed standard variance 10 nm/min.}
}
\label{fig.speed_distribution}
\end{figure}

\add{The model presented above with two competing fibril morphologies is further supported by the measurements of speeds at which the fibrils are growing (Fig.~\ref{fig.speed_distribution}). The speed distributions seem to have two peaks, whose magnitudes depend on the glucagon concentration. At low glucagon concentration the dominant peak is at \mbox{$\sim\!\!100\, \textrm{nm/min}$}, which probably corresponds to the growing speed of twisted fibrils composed of glucagon monomers. On the other hand, at large glucagon concentration the dominant peak is at \mbox{$\sim\!20$--$30 \,\textrm{nm/min}$}, which probably corresponds to the growing speed of non-twisted fibrils composed of glucagon trimers.
}

In conclusion, we present a monomer-trimer model for glucagon fibrillation and
compared it with our experimental data.  The model predicts a concentration
dependent growth probability, which we test experimentally at various glucagon
concentrations by analyzing the distributions of growth and stasis duration.
Our model captures the short time behavior of growth and pause durations and
reproduce the experimentally observed growth probability well.  The stop-go
kinetics observed requires two contrasting precursor states, one of which
elongates while the other one blocks.  Thus, the model might generically also
explain, e.g., fibril growth kinetics for $\beta$-lactoglobulin which exists in
a monomer-dimer equilibrium, where only the monomer is capable of elongating
fibrils (via a partially unfolded state) \cite{hamada02}.

\begin{acknowledgments}
This work was supported by the Lundbeck Foundation (BioNET2),
the Danish National Research Foundation through the Center for Models of Life and
the KU Center of Excellence (MolPhysX).
We are grateful to Novo Nordisk A/S for providing glucagon samples.
\end{acknowledgments}
\bibliography{growthmodel_lib}
\end{document}